\documentclass{article}
\usepackage{lipsum}
\usepackage{amsmath}
\usepackage{gensymb}
\usepackage{graphicx,booktabs,array}
\usepackage{blindtext}
\usepackage{titlesec}
\usepackage{slashed}
\usepackage{epsfig}
\usepackage{comment}
\usepackage{cancel}
\usepackage{bbm}
\usepackage{array}
\usepackage[table]{xcolor}
\usepackage{bigints}
\usepackage{booktabs}
\usepackage{color}
\usepackage{dsfont}
\usepackage{tensor}
\usepackage{float}
\usepackage{framed}
\usepackage{graphicx}
\usepackage{indentfirst}
\usepackage{mathrsfs}
\usepackage{multirow}
\usepackage{subdepth}
\usepackage[dotinlabels]{titletoc}
\usepackage{wrapfig}
\usepackage[all]{xy}
\usepackage[vcentermath]{youngtab}
\usepackage{relsize}
\usepackage{hyperref}
\numberwithin{equation}{section}

\usepackage[utf8]{inputenc}
\usepackage{csquotes}
\usepackage{slashed}
\usepackage{amsmath}
\usepackage{amsfonts}
\usepackage{amssymb}

\begin{document}

\begin{titlepage}
\unitlength = 1mm
\ \\
\vskip 3cm
\begin{center}
 
{\huge{\textsc{Necessitating Spacetime
}}}

\vspace{1.25cm}
Kyle Singh,$^{\mathsection}\footnote{kyle.singh.gr@dartmouth.edu}
$Jenna Van Dyke$^{*}$\footnote{jennavandyke@ucsb.edu}

\vspace{.5cm}

$^\mathsection${\it  \small{Guarini School of Graduate and Advanced Studies,
64 College Street,\\
Anonymous Hall Suite 102
Hanover,
New Hampshire, 03755}}

$^*${\it\small{University of California, Santa Barbara, Department of Physics, \\ Broida Hall, Santa Barbara, CA 93106} }

\vspace{0.8cm}

\providecommand{\keywords}[1]
{
  \small	
  \textbf{Keywords:} #1
}

\begin{abstract} 

We investigate structure that describes physical data in gravitational systems that is, to one degree or another, independent of the metric and affine structure. We dub such structure surplus structure and seek to incorporate it into our ontological commitments. An emphasis is placed on those structures which are required to constrain our models motivated by physical data. We look at the fall-offs of the gauge fields in asymptotically flat spacetimes, the computation of entanglement entropies in the AdS/CFT correspondence, and the addition of arbitrary parameters which modify the causal structure of spacetime in Kruskal coordinates. We also present a historical overview of the understanding of singularities in classical gravitation. Primary sources are turned to here. A toy cosmological model is also explored in which global topology is determined by an additional function which we prescribe as a dynamical $\Lambda$. Such surplus structure is classified using the language of intermediaries and positioned within the system of a shifting-scale ontology as introduced in prior work.

\end{abstract}

\vspace{1.0cm}

\end{center}

\end{titlepage}

\pagestyle{empty}
\pagestyle{plain}

\def\vx{{\vec x}}
\def\p{\partial}
\def\po{$\cal P_O$}
\def\i{{\rm initial}}
\def\f{{\rm final}}

\pagenumbering{arabic}

\newpage

\tableofcontents

\section{Introduction}
Let us recall the statement characterized by Newton's shell theorem wherein matter inside a spherical shell has no gravitational pull on it. We can show that this also holds in GR; namely that the metric within the shell is simply the Minkowksi metric for a vacuum. The metric for a static spherical spacetime can be written as follows
\begin{equation}
ds^2=-e^{2\Phi}dt^2+e^{2\Lambda}dr^2+r^2d\Omega^2
\end{equation}	  
Setting the $G_{tt}$ component of the Einstein tensor to 0, one finds the following expression
\begin{equation}
G_{tt}=0=\frac{d}{dr}\left [ \left ( 1-e^{-2\Lambda} \right ) \right ]
\end{equation}
which integrates to 
\begin{equation}
e^{2\Lambda}=\frac{r}{r+C_1}
\end{equation}
Using this result, one finds the following expression from the $rr$ component 
\begin{equation}
\Phi'=\frac{1}{2r^2}e^{2\Lambda}C_1
\end{equation}
solving for $\Phi$ yields the following
\begin{equation}
\Phi(r)=\frac{C_1}{2}\displaystyle\int\frac{dr}{r^2+rC_1}=\frac{1}{2}\text{ln}\left ( \frac{r}{r+C_1} \right )+C_2
\end{equation}
Rearranging terms gives us the following expression
\begin{equation}
e^{2\Phi(r)}=C_3\left ( \frac{r}{C_1+r} \right )
\end{equation}
With all of this in hand, we assume a non-degenerate metric and impose a boundary condition at $r=0$. Eq. (1.6) sets $C_1=0$. Moreover, we now know that $g_{tt}=-C_3$ given the conditions above. Absorbing $C_3$ into $t$ gives us the desired Minkowski metric in spherical coordinates
\begin{equation}
ds^2=-dt^2+dr^2+r^2(d\theta^2+sin^2\theta d\phi^2)
\end{equation}
The above calculation demonstrates a simple redux on the standard hole argument in which the ``hole"-region is the recovered Minkowski space. \footnote{Of course, this is a loose equivalence brought up to carry out the succeeding argument}

The above is an example of the way in which physical requirements, and corresponding physical reasoning, could in fact fix the metric structure of two distinct regions of spacetime independent of any diffeomorphism. Indeed, the transition here is not one that occurs via a symmetry transformation, rather it occurs wholly in virtue of the physical requirements set forth by classical gravitational constraints. Such considerations of structure, which will be relevant to a refinement of our ontological commitments in the context of classical gravitation which go beyond affine structure, were first brought up by Curiel. He writes in [1]
 \begin{displayquote}
In sum, the hole argument has no bearing on whether existence should be attributed to spacetime points independent of metrical structure. The diffeomorphic freedom in the presentation of relativistic spacetimes does not \emph{ipso facto} require philosophical elucidation, for it in no way prevents us from investigating what is of true physical significance... I think the most unproblematic and uncontroversial fact about diffeomorphic freedom is that it embodies an inevitable arbitrariness in the mathematical apparatus the theory uses to model physical systems: the choice of the presentation of the spacetime manifold and metric one uses to model a physical system is fixed only up to diffeomorphism. (Curiel, 7-8)
\end{displayquote}
Maudlin shares a similar sentiment in [2] when he writes that the ambiguity in physical states via diffeomorphism invariance leads to ``radical indeterminism" (Maudlin, 85). He expands
\begin{displayquote}
the problem arises as soon as one introduces names for substances and further assumes that the substances have all of their properties only accidentally. For a world made up of bare particulars and accidents automatically falls prey to a kind of permutation argument: permute some of the bare particulars, have them exchange all of the properties with which they have been clothed, and the result is a distinct yet observationally identical state of affairs. (Maudlin, 85)
\end{displayquote}
Maudlin further pursues the notion that if one restricts themselves to manifold substantivalism, they forgo a variety of a spatiotemporal characteristics of GR such as the light-cone structure which preserves causality. Moreover, it seems as though, in the spirit of Curiel, when we phrase the ontological question purely in terms of the metric, or the manifold, or the set of spacetime points, we are forced to leave out those structures and physical constraints which can not be derived, in some way or another, from those entities directly.

The question then becomes precise if we ask ourselves what structures, which are independent of the metric in some sense, are indeed physically relevant. Such structures \emph{necessitate} spacetime in the sense that they are ontological to the description of physical phenomena. Although there is ample debate on the ontological status of spacetime itself, what we argue is that there is some sense in which the structure we will explore below is ontological in itself. In other words, such structure necessitates spacetime in the sense that it characterizes \emph{physical} data independent of the manifold or the metric. Still, this structure operates in the arena of phenomena where physical outcomes can be derived in a gravitational context. Of course, our notions of independence will need to be precisely defined. Indeed, through each piece of surplus structure explored below, we constantly redefine our notions of independence. In many ways, this is the central question. Addressing it is the end result of classifying such structures at large. In [9], we attempted to do this in the context of gauge symmetries. In this work a shifting-scale ontology was introduced in which ontological commitments are altered by initial conditions of a physical system. We classify all such surplus structures as intermediary in the spirit of the previous discussion on the status of gauge symmetry and its role in physical theory. As of now, this is how we choose to navigate the tension between the invocation of such entities as ontological and the fact that they need to be put in to the physical apparatus ``by hand," so to speak. \footnote{Curiel attempts to do this, for example, by invoking auxiliary scalar field which are responsible for containing relevant physical data, ``Points in a spacetime manifold have existence independent of metrical structure if the manifold can be constructed from a family of scalar fields, the values of which can be empirically determined without knowledge of metrical structure" (Curiel, 18). Our approach is somewhat different as we are, again, exploring structures which may influence the manifold as it stands and in whatever way it is to be categorized. }\footnote{We put the emphasis here in the spirit of Curiel in [1] when he makes an appeal to physical consequence over mathematical abstraction for its own sake.}

Curiel explores this at some length when he invokes the Killing equation 
that generates isometries in GR
\begin{equation}
    \nabla_{(a}\xi_{b)}=0
\end{equation}
Curiel makes the arguments that such fields are, in certain contexts, physical and are indeed independent of metrical structure. In this work, we wish to explore a broader class of such structures and evaluate the ontological status of them in greater detail and within the framework constructed in discussions of the gauge principle. 

Such structures are, again, motivated by physical reasoning. In bringing them up, the issue of whether the essence of spacetime consists of a manifold or, rather the spacetime points, is separate from the appropriate categorization of such mechanisms that are independent from the metric or affine structure and yet essential in some way to physical processes. Instead, in trying to categorize entities that are  distinct from the two categories relevant to the standard relationist vs. substantivalist debate, we wish to ascribe to them a space in our ontological commitments. These structures will influence the manifold and perhaps even interact with the metric, regardless of what status we ascribe to them, but their construction can be entirely independent of the metric or affine structure in some instances. We navigate such negotiations of independence as it seems apt. We investigate asymptotic fall-offs in Bondi coordinates, the calculation of holographic entropy in the AdS/CFT correspondence, the addition of an arbitrary parameter that affects Black Hole topology in the Kruskal spacetime, and a toy cosmological model where $\Lambda$ is a function of the spacetime coordinates itself and attempt to classify each individually as follows. 
\section{Asymptotically Flat Spacetimes}
An early example of the study of asymptotic symmetries\footnote{Note that here, we refer to asymptotic symmetries as the study of conserved charges or nontrivial exact symmetries in any system with a boundary region} in GR was conducted by Bondi, van der Burg, Metzner, and Sachs [13,14]. Their original goal of recovering the Poincaré group from SR as the symmetry group of asymptotically flat spacetimes resulted instead in the discovery of the BMS group, that is infinitely dimensional and encompasses the full set of symmetry transformations at the boundary. An example of such transformations are the super-translations and super-rotations which are found at past and future null infinity. An important piece of this analysis involves the determination of fall-off conditions are imposed for the fields at $\mathcal{I}$.\footnote{Here $\mathcal{I}$ refers to either past or future null infinity. The notation comes from the standard Penrose diagram}Let us investigate one way fall-offs can be determined in the context of electromagnetism. Recall that the Maxwell equations can be written in the following way, which is manifestly Lorentz covariant
\begin{equation}
\nabla_\mu F^{\mu\nu}=j^\nu
\end{equation}
In curvilinear coordinates the LHS takes on the following form
\begin{equation}
\nabla_\mu F^\mu_\nu=\partial_\mu F^\mu_\nu+\Gamma^\mu_{\mu\rho}F_\nu^\rho-\Gamma^\rho_{\mu\nu}F_\mu^\rho
\end{equation}
Let us also work in Lorenz gauge where
\begin{equation}
\nabla_\mu A^\mu=0=\frac{1}{\sqrt{g}}\partial_\mu\left ( \sqrt{g}A^\mu \right )=\frac{1}{\sqrt{g}}\partial_\mu\left ( \sqrt{g}g^{\mu\nu}A^\mu \right )
\end{equation}
We now write the line element in retarded Bondi coordinates \footnote{Here, note that $\gamma_{z\bar{z}} =\frac{2}{(1+z\bar{z})^2}$ and is the metric on the 2-sphere}
\begin{equation}
ds^2=\eta_{\mu\nu}dx^\mu dx^\nu=-du^2-2dudr+2r^2\gamma_{z\bar{z}}dzd\bar{z}
\end{equation}
The Maxwell equation in Lorentz gauge becomes 
\begin{equation} 
\begin{split}
\frac{1}{r^2\gamma_{z\bar{z}}}\partial_u\left ( r^2\gamma_{z\bar{z}}(-1)A_r \right )+\frac{1}{r^2\gamma_{z\bar{z}}}\partial_r\left ( r^2\gamma_{z\bar{z}}(-1)A_u \right )+
\frac{1}{r^2\gamma_{z\bar{z}}}\partial_r\left ( r^2\gamma_{z\bar{z}}A_r \right )\\+\frac{1}{r^2\gamma_{z\bar{z}}}\partial_z\left ( \frac{1}{r^2\gamma_{z\bar{z}}}r^2\gamma_{z\bar{z}}A_{\bar{z}} \right )+\frac{1}{r^2\gamma_{z\bar{z}}}\partial_{\bar{z}}\left ( \frac{1}{r^2\gamma_{z\bar{z}}}r^2\gamma_{z\bar{z}}A_{z} \right )=0
\end{split}
\end{equation}
We can rewrite this as follows
\begin{equation} 
-\partial_u(r^2A_r)-\partial_r(r^2A_u)+\partial_r(r^2A_r)+\gamma^{z\bar{z}}(\partial_z A_{\bar{z}}+\partial_{\bar{z}}A_z)=0
\end{equation} 
Then the large $r$ fall-offs are 
\begin{subequations}
\begin{equation} 
A_u\sim\frac{1}{r} 
\end{equation} 
\begin{equation}
A_r\sim \frac{1}{r^2}
\end{equation}
\begin{equation}
A_z\sim \mathcal{O}(r^0)
\end{equation}
\end{subequations}
One can now expand the fields for any particular component in a large-$r$ expansion
\begin{equation} 
A_z(r)=\sum_{n=0}\frac{A_z^{(n)}}{r^n}
\end{equation} 
and can obtain the following expressions
\begin{subequations}
\begin{equation} 
A_r=\frac{A_r^{(2)}}{r^2}+\frac{A_r^{(3)}}{r^3}+...\end{equation} 
\begin{equation}
A_u=\frac{A_u^{(1)}}{r}+\frac{A_u^{(2)}}{r^2}+...\end{equation}
\begin{equation}
A_z=A_z^{(0)}+\frac{A_z^{(1)}}{r}+...
\end{equation}
\end{subequations}
Then, at leading order, one obtains the following Maxwell equation in Lorenz gauge written for large-$r$ 
\begin{equation} 
\partial_uA^{(2)}_r-A^{(1)}_u+\gamma^{z\bar{z}}\left ( \partial_zA^{(0)}_{\bar{z}}+\partial_{\bar{z}} A_z^{(0)}\right )=0
\end{equation}
We can also conduct a similar analysis for the Maxwell equations in the presence of a sourced current.
\begin{equation} 
\nabla_\mu F^{\mu\nu}=j^\nu \end{equation}
Where the current falls off as follows
\begin{equation}
j_u\sim \frac{1}{r^2}
\end{equation}
We now wish to write this equation in the Bondi coordinates. We consider the case where $\nu=0=u$ and calculate explicitly one component of the field strength to show how it falls off at large-$r$. 
\begin{equation}
\begin{split}
    F_{ru}=\partial_r A_u -\partial_uA_r&=\partial_r\left ( \frac{1}{r}A_u^{(1)}+\frac{1}{r^2}A_u^{(2)}+... \right )-\partial_u\left ( \frac{1}{r^2}A_r^{(2)}+\frac{1}{r^3}A_r^{(3)}+... \right )\\ &=\left (  -\frac{1}{r^2}A_u^{(2)}+\frac{2}{r^3}A_u^{(3)}+...\right )-\left ( \frac{1}{r^2}A_r^{(2)}-\frac{1}{r^3}A_r^{(3)} \right )
    \end{split}
\end{equation}
This implies that this particular component of the field strength falls off like $\frac{1}{r^2}$.
This leads to the following Maxwell equation 
\begin{equation}
\partial_u F_{ru}^{(2)}+(\partial_zF_{z\bar{z}}+\partial_{\bar{z}}F_{uz})\gamma^{z\bar{z}}=j_u^{(2)}
\end{equation}
The  fall-offs that arise in the above asymptotic expansion  are not determined by a particular physical theory but are additional requirements needed in order to parameterize the space of solutions that may or may not have physical meaning at large-$r$. Indeed, the above analysis yields structure that is completely independent from any notions of curvature or a given spacetime metric. 

It must be noted, and is perhaps already been inferred by the reader at this point, that there is a certain amount of ambiguity in the determination of large-$r$ fall-offs. Indeed, we have assumed that the corresponding expansions of the fields hold. What fall-off conditions we choose depend on first and foremost the class of solutions we are interested in exploring in a given theory. For example, if we want to find solutions to Einstein's equations which satisfy asymptotic flatness at null infinity, then we can in principle look for the \emph{most general} fall-off conditions that satisfy asymptotic flatness [12].  This solely comes from the condition of asymptotic flatness, which in principle is a limiting case of any spacetime. It is clear that this first imposition of fall-offs is determined by curvature and a particular background geometry. Indeed, if we assume the most general fall-offs, we are still left with the issue raised in the preceding sections of whether or not we transition to a unique physical state when an analysis is done in with a new metric structure. 

For such surplus structure to be ontological in some sense, given the context of a shifting scale ontology, it is crucial, as argued in the preceding section, that such structure is as \emph{independent} of the curvature as possible. Another way such fall-offs can be determined is with the imposition of additional constraints which may be physically motivated. For example, based on our experience with, say classical gravitational scattering, we can demand that the radiative field should fall-off at a certain rate as we go  to the future and past spheres of null infinity. Of course asymptotic flatness will still itself  put a certain constraint on the fall-offs of radiative data as $u \rightarrow \pm \mathcal{I}$, but fall-offs that are physically relevant and that are more stringent must be put in by hand. \footnote{One example of this in the study of asymptotic symmetries is the relaxation of late-$u$ fall-offs of the radiative data in order to obtain a larger solution space. It is not always clear that once we relax fall-off conditions on radiative data in this way, how asymptotic flatness affected. This has to be checked depending on what conditions one chooses and has been done in a variety of contexts.}\footnote{The calculational details can be found in the aforementioned sources.}

That the fall-offs which are physical must be put in by hand, in some cases, means that they are, at some level, independent of the affine structure.  They require a piece of our ontological commitments because they reveal aspects of the physical structure that can not be derived from the affine architecture. Moreover, since Bondi coordinates are, in principle the geometric limit of any conceivable spacetime, they are ubiquitous. Granted, this notion is weaker due to our notions of diffeomorphism invariance. Independent of whether or not a spacetime constitutes a new physical state, there is a sense in which all spacetime are related by diffeomorphisms. However, again, these fall-offs can not be obtained via such a transformation as we have been alluding too. Thus, they come to represent our first example of surplus structure as defined in the preceding section.

\section{Entanglement Entropies in AdS/CFT} 
Maldacena's discovery of the AdS/CFT correspondence [7] presents the sharpest example of the so-called holographic principle which purports that particular gravitational theories in the bulk, a certain $d$-volume in space, can be described equally by a quantum theory on the the boundary in lower dimensions. We explore this correspondence in the context of entanglement entropies. Ryu and Takayanagi conjectured [11] that the area of a minimal surface in the AdS bulk is directly related to the entanglement entropy\footnote{Recall that the Von Neumann entropy is given by the following formula \begin{equation}
    S=-\text{tr}\rho\text{log}\rho
\end{equation}}  between two subsystems  $A$ and $B$ in a corresponding CFT by the following formula 
\begin{equation}
S_A=\frac{\gamma_A}{4G_N}
\end{equation}
Here $\gamma$ is the minimal surface and $G_N$ is Newton's gravitational constant. More precisely, one needs to restrict the AdS spacetime to a Poincaré patch and compute the entropy between two subsystems on a corresponding conformal boundary.\footnote{For a full review of the correspondence, one can refer to [8]} We compute the corresponding entropy in $\text{AdS}_4$/$\text{CFT}_3$ as follows. Let us begin with the spatial part of the metric restricted to a Poincaré patch in $\text{AdS}_4$
\begin{equation}
    dl^2=\frac{L^2}{z^2}\left( dr^2+r^2d\theta^2+dz^2 \right)
\end{equation}
The minimal surface we are interested in is a geodesic centered at the conformal boundary with radius $R$
\begin{equation}
    \left( x-x_0\right)^2+\left( y-y_0\right)^2+z^2=R^2
\end{equation}
Note that the above is independent of the $x$ and $y$ coordinates respectively. Using this translational invariance we differentiate the above and recast the metric in cylindrical coordinates to obtain the following expression
\begin{equation}
    d\sigma^2=\frac{L^2R^2}{\left(R^2-r^2\right)^2}dr^2+\frac{L^2R^2}{R^2-r^2}d\theta^2
\end{equation}
We can now compute the area as follows
\begin{equation}
\gamma_A=\displaystyle\int_0^{2\pi}d\theta\displaystyle\int_0^Rdr\sqrt{h}=\frac{2\pi L^2R}{\epsilon}-2\pi L^2
\end{equation}
Here $h$ is the induced metric given by eq. (3.5) and $\epsilon$ is some regulator in the spatial coordinates introduced to ensure that the integral converges. The entropy can then be written as follows
\begin{equation}
S_A=\frac{\pi L^2R}{2G_N}\frac{1}{\epsilon}-\frac{\pi L^2}{2G_N}   
\end{equation}
A similar calculation yields the following result in $\text{AdS}_3$/$\text{CFT}_2$
\begin{equation}
S_A=\frac{c}{3}\text{ln}\frac{L}{\epsilon} 
\end{equation}
Both results yield exact equivalences with what is obtained in the corresponding CFTs. In order to categorize such observables, and the context in which they arise, we must first state that we take the duality between AdS spacetimes and CFTs as representing a one-to-one map between, in this case, thermodynamic quantities. This needs to be spelled out since, as is well known, the AdS/CFT correspondence is treated still as a conjecture, although a myriad of computations, as the one performed above, suggest that it holds in the strongest sense. Indeed, some have pointed out that there must be some order in perturbation theory in which this duality breaks down, which results in the fact that spacetime structure must be emergent in some sense and  nonexistent at the quantum level. This was first explored heuristically in [22]. For now, we take the former approach and treat these two theories as true one-to-one maps. 

In this frame, we have an intermediary structure in the strongest sense, namely in the form of an \emph{entirely different} theory. The CFT on the boundary is a quantum theory completely separate from the bulk theory. One can then infer that any relevant physical quantities derived hold from the boundary to the bulk. This example can be seen as the simplest case of surplus structure and maintains that, perhaps, we yield it our strongest ontological weight since we are operating in a theory which, in its entirety, is the intermediary structure. We can bypass questions of manifolds and metrics if we treat this boundary theory as ontological. Of course, this is a crude statement given that much of the physical structure of GR cannot currently be captured by such quantum theories, although it is conceivable that some day more phenomena may find itself being expressed in this domain. If so, we add this theory to the list of relevant entities, and hold again that its categorization is strong and straightforward, unlike what we will explore next.

\section{Historiography of Singularities} 
We now turn our attention to a discussion of Black Holes and their singularity structure. We present a historiographic analysis here to provide insight into the ambiguity of singularities over the course of \emph{scientific practice} in itself. 

In formulating GR, Albert Einstein required that his theory satisfy three criteria: it must approximate Newton's theory, be covariant under coordinate transformations, and satisfy the equation of the determinant
\begin{equation}
    |g_{\mu\nu}|=-1
\end{equation}
which allows the equations of the theory to be greatly simplified. In their final form [19], his equations read
\begin{equation}
    \sum_{\alpha\beta}\frac{\partial^2g^{\alpha\beta}}{\partial x_\alpha\partial x_\beta}-\kappa(T+t)=0
\end{equation}
where $T$ and $t$ are matter and energy components respectively. Following Einstein's work, Karl Schwarzschild [18] endeavored to model the gravitational field of a mass point, in this case around a static, spherically-symmetry mass.

 \begin{displayquote}
Das Problem ist, ein Linienelement mit solchen Koeffizienten
ausfindig zu machen, daß die Feldgleichungen, die Determinantengleichung und diese vier Forderungen erfüllt werden. (Schwarzschild, 2)
\end{displayquote}

\begin{displayquote}
The problem is to find out a line element with coefficients such that the field equations, the equation of the determinant and these four requirements are satisfied.
\end{displayquote}
The four requirements he refers to here are time-independence, diagonality, rotational symmetry, and asymptotic flatness. The result is the following line element
\begin{equation}
    ds^2=(1-\alpha/R)dt^2-\frac{dR^2}{1-\alpha/R}-R^2(d\theta^2+\sin^2\theta d\phi^2)
\end{equation}
where $R=(r^3+\alpha^3)^{1/3}$, the ``Unstetigkeit" (discontinuity) is at $r=(\alpha^3-\rho)^{1/3}$, and the parameter $\rho$ may offset the discontinuity from the origin. There are thus two singularities of the Schwarzschild line element, at $R=0$ and $R=\alpha$. Of course we know this latter discontinuity to be a false coordinate singularity; in reality there is only the event horizon of the Black Hole at $\rho=\alpha^3$ and $R=\alpha=2m$. As we will see, the real physical structure of the Schwarzschild geometry was gradually peeled away over several decades. In his 1916 article \textit{Die Grundlagen der Physik} (\textit{The Foundations of Physics}) [23], David Hilbert produces one of the first definitions of a non-singular (``regulär") spacetime.

\begin{displayquote}
    Dabei nenne ich eine Maßbestimmung oder ein Gravitationsfeld $g_{\mu\nu}$ an einer Stelle regulär, wenn es möglich ist, durch umkehrbar eindeutige Transformation ein solches Koordinatensystem einzuführen, daß für dieses die entsprechenden Funktionen $g'_{\mu\nu}$ an jener Stelle regulär d. h. in ihr und in ihrer Umgebung stetig und beliebig oft differenzierbar sind und eine von Null verschiedene Determinante $g'$ haben. (Hilbert, 70-71)
\end{displayquote}

\begin{displayquote}
    By that I mean that a line element or a gravitational field $g_{\mu\nu}$ is regular at a point if it is possible to introduce by a reversible, one-one transformation a coordinate system, such that in this coordinate system the corresponding functions $g'_{\mu\nu}$ are regular at that point, i.e., they are continuous and arbitrarily differentiable at the point and in a neighborhood of the point, and the determinant $g'$ is different from 0.
\end{displayquote}
There are two main aspects of Hilbert's definition to be discussed here. First, his identification of point singularities would, in some contexts, be considered ill-defined. The reason for this is that singular structure indicates that the spacetime is somehow not mathematically well-defined, and thus one cannot reliably define the precise location of a point singularity. Einstein and Rosen [24], for one, have expressed a similar sentiment, writing, ``For a singularity brings so much arbitrariness into the theory that it actually nullifies its laws" (Einstein and Rosen, 73). Instead, one might say that the spacetime as a whole is singular. Second, the requirement for continuity alludes to the idea of geodesic incompleteness, perhaps the most characteristic quality of a singular spacetime and its most simple definition. On a final note, the distinction between real singularities and coordinate singularities has also at this point yet to be made. Arthur Stanley Eddington's 1923 text, \textit{The Mathematical Theory of Relativity} [25], provides an early analysis of singularities. He inspects the Schwarzschild spacetime (4.3) by comparing it to de Sitter's maximally-extended spherical world, whose line element reads
\begin{equation}
    ds^2=\gamma dt^2-\gamma^{-1}dr^2-r^2(d\theta^2+\sin^2\theta d\phi^2)
\end{equation}
where $\gamma=1-\frac{1}{3}\lambda r^2=1-\alpha/R$. He identifies the analogous singularities of each spacetime at $r=\sqrt{3/\lambda}$ and $R=\alpha$, which brings into question whether de Sitter spacetime contains a ``mass-horizon" like Schwarzschild does.

\begin{displayquote}
A singularity of $ds^2$ does not necessarily indicate material particles, for we can introduce or remove such singularities by making transformations of coordinates. It is impossible to know whether to blame the world-structure or the inappropriateness of the coordinate-system. (Eddington, 279)
\end{displayquote}
With this, Eddington makes the landmark distinction between real and coordinate singularities. Although it was thought impossible to figure out which type one is dealing with, it is intuitive to try and find out by computing some physical quantity invariant to the coordinate system. Upon calculating the gravitational flux in de Sitter coordinates to be $4\pi r^2(-\delta y-2\delta\gamma/r)dt$ and substituting $\gamma$, he finds that it vanishes for all values of $r$, showing that the spacetime is indeed empty. Hence he concludes, ``I believe then that the mass-horizon is merely an illusion of the observer at the origin, and that it continually recedes as we move towards it" (Eddington, 280). One can show that the Schwarzschild singularity is entirely illusory by performing a similar procedure with any coordinate-invariant quantity. Eddington's own coordinate system [26] was the first to resolve this false singularity, although he did not initially seem aware of this. His actual motivation was to compare the equations of GR with Whitehead's gravitational theory, which at the time was considered a plausible alternative. In order to show a certain equivalence between each theory's field equations, he devised new a coordinate system for the Schwarzschild metric in which radial null geodesics are unity. After performing the transformation $t_1=t-2m\log(R-m)$, he obtains the line element
\begin{equation}
    ds^2=dt_1^2-dr^2-r^2(d\theta^2+\sin^2\theta d\phi^2)-(2m/R)(dt_1-dr)^2
\end{equation}
It is clear that the singularity at $R=2m$ has been resolved, but Eddington makes no comment on this in his work. His coordinates were subsequently lost until David Finkelstein rediscovered them [27] in 1958. It was not until 1933 that Georges Lemaitre [28] explicitly described for the first time the singularity's illusory nature, using his own transformation of coordinates. Before Martin Kruskal published the maximally-extended solution, Christian Fronsdal [29]  recovered a similar result via a six-dimensional embedding of the Schwarzschild solution. He obtained the following surface
\begin{equation}
    \begin{split}
        -Z_1^2+Z_2^2&=4(1-1/r) \\
        Z_3&=\int dr(r^2+r+1)r^{-3/2} \\
        Z_4^2+Z_5^2+Z_6^2&=r^2
    \end{split}
\end{equation}
The line element is then
\begin{equation}
\begin{split}
    ds^2&=(1-1/r)dt^2-(1-1/r)^{-1}dr^2\\
    &=dZ_1^2-dZ_2^2-dZ_3^2
\end{split}
\end{equation}
where the radius of the event horizon has been set $r=2m=1$ and $d\theta=d\phi=0$. However, this coordinate system retains the false singularity at $r=1$ and Fronsdal's extension is thus not fully maximal. Still, he is able to recover the two interior and two exterior regions of the spacetime.



In 1960, Kruskal [19] and Szekeres [30] independently present the complete maximal extension. Kruskal is motivated by a spherically-symmetric spacetime in which radial null geodesics have a slope of $\pm1$. This yields
\begin{equation}
    ds^2=f^2(-dv^2+du^2)+r^2d\omega^2
\end{equation}
where $f^2=(32m^3/r)e^{-r/2m}$ completes the successful transformation from Schwarzschild coordinates:
\begin{equation}
    \begin{split}
        u=\left(\frac{r}{2m}-1\right)^{1/2}e^{\frac{r}{4m}}\cosh\left(\frac{t}{4m}\right) \\
        v=\left(\frac{r}{2m}-1\right)^{1/2}e^{\frac{r}{4m}}\sinh\left(\frac{t}{4m}\right)
    \end{split}
\end{equation}
We also have the following useful relations
\begin{equation}
    u^2-v^2=\left(\frac{r}{2m}-1\right)e^{r/2m}
\end{equation}
\begin{equation}
    \frac{v}{u}=\tanh\left(\frac{t}{4m}\right)
\end{equation}
Let us then plot lines of constant $r$ and $t$ in the $u$-$v$ plane, where (4.11) is the slope. This yields the Kruskal diagram, which reveals all four regions of the Schwarzschild geometry. The original Schwarzschild coordinates (4.3) may only access the Black Hole exterior, while Eddington-Finkelstein coordinates (4.5) recover the interior and exterior. With Kruskal coordinates (4.8), we may now also access the theoretical White Hole interior and exterior. 
\begin{displayquote}
    It is remarkable that it presents just such a ``bridge" between two otherwise Euclidean spaces as Einstein and Rosen sought to obtain by modifying the field equations. It may also be interpreted as describing the ``throat of a wormhole" in the sense of Wheeler, connecting two distant regions in \textit{one} Euclidean space--in the limit when this separation of the wormhole mouths is very large compared to the circumference of the throat. (Kruskal, 1743)
\end{displayquote}
Einstein and Rosen [24] first spoke of this ``bridge" in the context of the Black Hole interior and exterior, and it can now be further extended to the global geometry. We are also interested in this idea of a wormhole ``throat." Kruskal's final note on this reads,
\begin{displayquote}
in effect the throat ``pinches off" the light ray before it can get through. This pinch-off effect presents fundamental issues of principle which require further investigation. (Kruskal, 1745)
\end{displayquote}
In the following section we too address this fact by inspecting an embedding surface of the Kruskal solution.

\section{Exploring Parameter Space}
For constant $V$, and by exploiting the inherent spherical symmetry, the Kruskal metric can be written as 
\begin{equation}
    d\Sigma^2=\frac{32M^2}{r}e^{-(r/2M)}du^2+r^2d\phi^2
\end{equation}
As is commonly done, we can write $U$ as a function of $r$ for a given $V$ as follows
\begin{equation}
    U^2(r,V)=\left ( \frac{2}{2M}-1 \right )e^{r/2M}+V^2
\end{equation}
Then, the metric becomes
\begin{equation}
 d\Sigma^2=\frac{re^{r/2M}}{2MU^2}dr^2+r^2d\phi^2
\end{equation}
We now wish to create an embedding surface by obtaining three functions in cylindrical coordinates in flat space by locating the point $(r,\phi)$ on the surface. Letting $\rho=r$ and $\psi=\phi$, one finds 
\begin{equation}
    \left (  \frac{dz}{dr}^2\right )\equiv A^2(r,V)\equiv\frac{re^{r/2M}}{2MU^2}dr^2-1
\end{equation}
And obtains the following form of the differential equation
\begin{equation}
    Z(r,V)=\displaystyle\int_{r_0}^r A(r^*,V)dr^*
\end{equation}
Numerically integrating the above expression gives us the following plots for a given parameterization of $V$.
\begin{figure}[h]
    \centering
    \includegraphics[width=5cm]{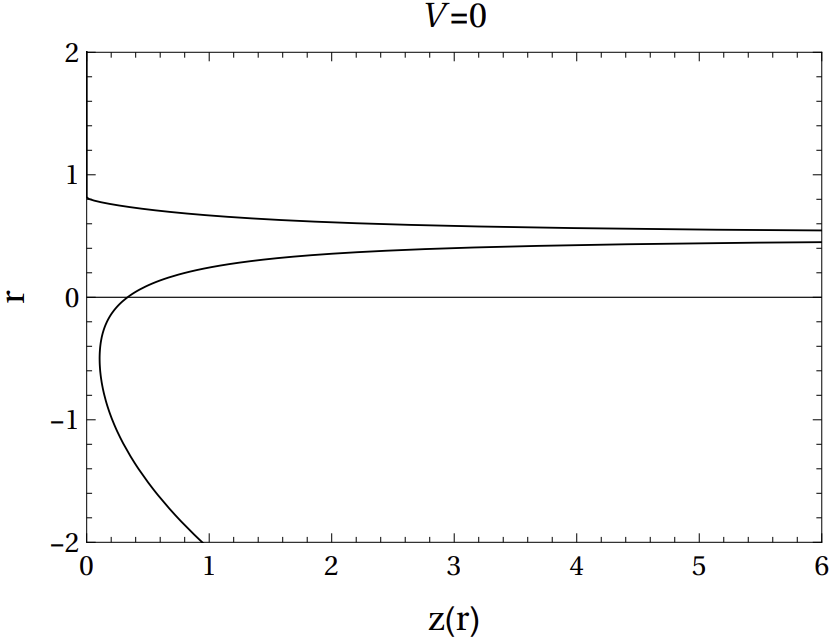}
    \includegraphics[width=5cm]{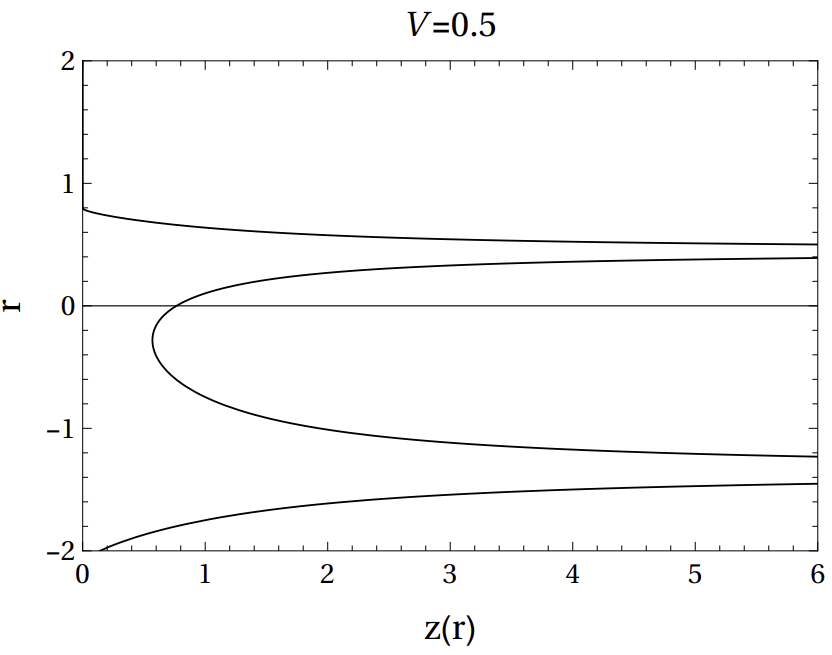}
    \includegraphics[width=5cm]{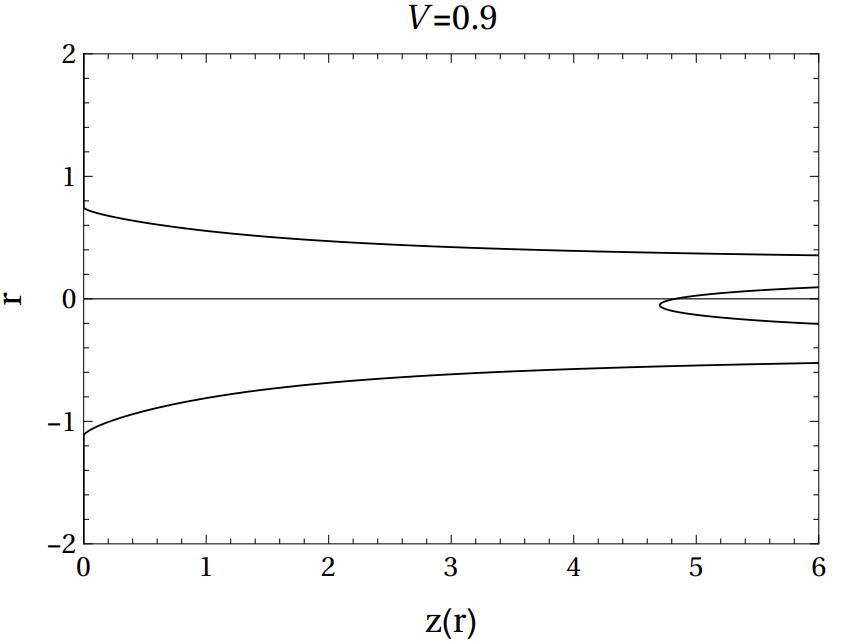}
    \includegraphics[width=5cm]{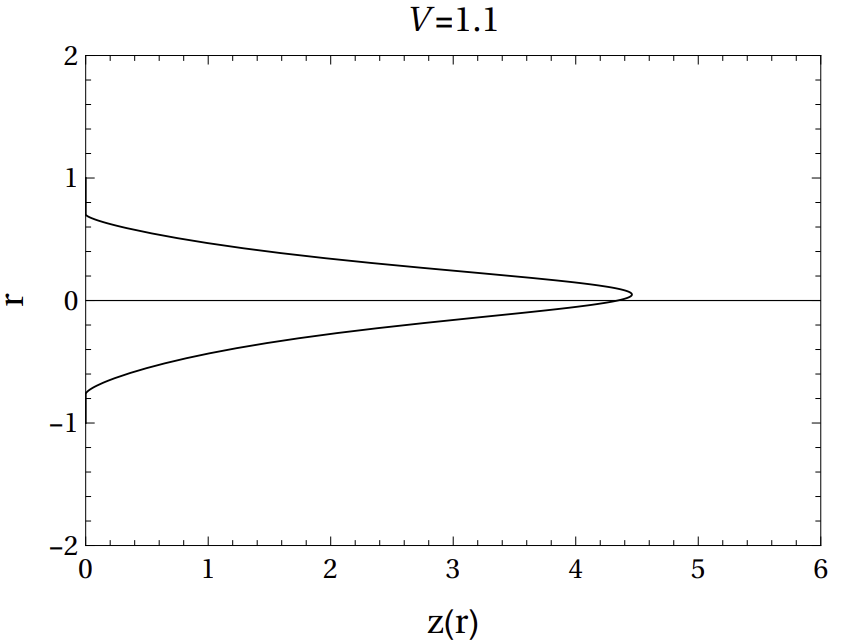}
    \caption{Kruskal embedding surfaces for varying parameterizations of $V$.}
    \label{fig:my_label}
\end{figure}

Notice that as $V$ approaches 1, the ``throat" separating the two causally disconnected spacetime regions begins to close. For all $V$ greater than 1, the two causally disconnected regions are now indistinguishable and the ``throat" has closed up entirely. In this case, the exterior structure introduced into the physical system is the arbitrary parameter $V$. Of the structures we have looked at thus far, this additional parameter seems to be most closely related to the prior metric structure that is given. After all, what physical meaning would it have if we were not working in the Kruskal coordinates? It is only given this particular physical meaning within the context of the embedding surface constructed. 

However, at the same time it is certainly straightforward to generalize the construction of such parameters to any spacetime. Indeed, one does this all the time in a variety of  physical models. Thus, such free parameters in general, must have a place in our ontological commitments.\footnote{Although it is clear, it must be said that their is a clear distinction between free parameters and fixed constants of which we do not make mention of here.} Any $V$ that we can construct that varies a set of coordinates in a particular spacetime, in which certain regimes of that $V$ alter the global geometry of that spacetime, is still strongly linked to the metric and is therefore weakly independent from the metric in the strongest sense. 

The ontology of such free parameters relies on the notion that they can always be introduced to toggle the physical structure of our system. They are somewhat independent from the metrical structure since they must always, again, be introduced by hand. That they must be introduced by hand qualifies them, in our language, as an intermediate structure. That they have a direct physical mapping onto the behavior of our system means that they are necessary since they reveal physical behavior about the system that could not be affirmed without their implementation. In this case the particular behavior would be the closing of the ``throat". We can separate such particular physical changes produced by the varying of the parameter and speak of the parameters, again in full generality by pointing to the fact that they can always be constructed and introduced within the context of a given metric. We add such parameters to our budding list of structures which are ontological and necessary to the physical mechanics of spacetime. 

\section{A Possible Inherent Feature of Spacetime} 

Two central problems in physical cosmology are an understanding of the Dark Matter and Dark Energy, both of which seem to be completely necessary for the evolution of our cosmology and play a key role in the $\Lambda$CDM model. Much has been said about the possibilities of these two entities and a variety of reviews are present in the literature. For the purposes of our investigation, we wish to focus on the DE as a possible dynamical background function inherent to the spacetime structure. In particular, in our toy model, $\Lambda$ is a function of the spatial coordinates itself. We investigate the corresponding cosmological evolution that arises from a particular choice of $\Lambda$. This follows the spirit of [4] in which Peebles and Ratra explore the notion of a time dependent $\Lambda$. We  begin with the following metric
\begin{equation}
ds^2=-e^{\Phi(\rho)}dt^2+a^2(t)d\rho^2+\rho^2d\phi^2+dz^2
\end{equation}
We can now write down the vacuum Einstein equation in the small-$\rho$ and large-$\rho$ regime. After making appropriate scaling arguments, We obtain the following expressions
\begin{subequations}
\begin{equation}
    2a\ddot{a}=\frac{e^{-\Phi(\rho)}\rho}{\Phi'}
\end{equation}
\begin{equation}
    \frac{\ddot{a}a}{\dot{a}}=e^{-\Phi(\rho)}\rho
\end{equation}
\end{subequations}
The above expressions have the following solutions for small-$\rho$ and large-$\rho$ respectively
\begin{subequations}
\begin{equation}
\text{erf}\left ( \frac{\pm\sqrt{\rho\text{ln}(t)+2C\text{exp}\left [ \kappa\text{sin}(\rho) \right ]\text{sin}(\rho)}}{\sqrt{\rho}} \right )=\frac{\sqrt{\rho}\text{exp}\left [ \frac{2C\text{exp}[\kappa\text{sin}(\rho)]\text{sin}(\rho)}{\rho}-\frac{\kappa\text{sin}(\rho)}{2} \right ]a(t)}{\sqrt{\pi}\sqrt{\text{sin}(\rho)}}+C_1
\end{equation}
\begin{equation}
\text{li}\left ( \exp\left [ {\frac{C\text{sin}(\kappa\rho)}{\rho}} \right ]t \right )=\rho\exp\left [ \frac{C\text{sin}(\kappa\rho)}{\rho}-\text{sin}(\kappa\rho) \right ]a(t)+C_1  
\end{equation}
\end{subequations}
where $\Phi(\rho)=\text{sin}(\kappa\rho)$. Now, let us consider the form of the stress energy tensor. For the proposes of simplicity in the toy model, we construct $T_{00}$ to only feature a dark energy component. Then the matrix takes on the following form
\begin{equation}
T_{\mu\nu}=\text{diag}(\kappa,0,0,0)
\end{equation}
We propose the following \emph{ansatz} for $T_{\mu\nu}$
\begin{equation}
T_{00}=\text{ln}\left ( 1+\alpha z \right )z^{-\beta-1}\equiv\Lambda(z)
\end{equation}
We can now impose the condition that the stress-energy momentum tensor is conserved, $\nabla_\mu T^{\mu\nu}=0$, to obtain the following constraint
\begin{equation}
H=\frac{\beta+1}{z}+\frac{1}{\text{ln}(1+\alpha z)(z+1)}
\end{equation}
Writing out the field equations and imposing appropriate scaling conditions, one finds the following expression
\begin{equation}
    H\rho+\frac{\partial\Phi}{\partial\rho}\frac{1}{2a^2\rho}-e^{-\Phi}\frac{\ddot{a}}{a}= \text{ln}(1+\alpha z)z^{-\beta-1}
\end{equation}
Using the result obtained from the conservation of the stress energy tensor, we can recast the above again in the small-$\rho$ and large-$\rho$ regimes
\begin{subequations}
\begin{equation}
\frac{\partial\Phi}{\partial\rho} \frac{1}{4a^2}-e^{-\Phi}\Omega^2=\Lambda(z)
\end{equation}
\begin{equation}
    \Omega \rho-e^{-\Phi}\Omega^2=\Lambda(z)
\end{equation}
\end{subequations}
Here, we let $\Omega=H$. We now have the set of equations which we utilize for numerical analysis (Table 1). We can analyze the behavior of this set-up in four regimes: small-$\rho$/large-$\rho$, in the presence of $\Lambda$/without the presence of $\Lambda$. Another way one could complete this analysis would be to utilize the analog of Stokes theorem in curved spacetime
\begin{equation}
\displaystyle\int_\mathcal{M}d^4x\sqrt{g}\nabla_\mu\left ( T^{\mu\nu} \xi_\nu\right )=\displaystyle\int_{\partial\mathcal{M}}d^3\sigma\sqrt{h}n_{\mu}T^{\mu\nu}\xi_\mu
\end{equation}
We can use this  to derive an expression for the boundary dynamics of our system. Recall, that here $h$ is the induced metric and $\xi$ a spacetime killing vector. Requiring a vanishing Lie derivative along a given vector field $K$,  $\pounds _kg_{\mu\nu}=0$, yields the following time-like killing vector field in $t$
\begin{equation}
\xi^\mu= \left ( -t\frac{\Phi'}{2},1,0,0,0 \right )
\end{equation}We compute the RHS as follows
\begin{equation}
\displaystyle\int_{\partial\mathcal{M}}d^3\sigma\sqrt{h}n_{\mu}T^{\mu\nu}\xi_\mu=\frac{1}{2}\displaystyle\int_0^\infty dta(t)t\displaystyle\int_0^\infty d\rho \rho\Phi'(\rho)\displaystyle\int_0^\infty dz\Lambda(z) 
\end{equation}
One finds the following solution for the integral over $z$
\begin{equation}
 \displaystyle\int_0^\infty T_{00}dz=\frac{\pi\alpha^\beta}{\beta\text{sin}(\pi\beta)}  
\end{equation}
for appropriate bounds in the free parameters $\alpha$ and $\beta$ such that the integral converges. We leave a full analysis of this method for future work and wish to note this as an alternative method to analyze the bulk-boundary dynamics in particular regime of $\rho$. One would, of course, have to specify a choice of $a(t)$. This could be derived using the expressions above. For now, let us analyze the results derived above for the dynamics in the specified regimes of interest.

\newcommand{\largerhowlambdaone}{\includegraphics[width=4cm]{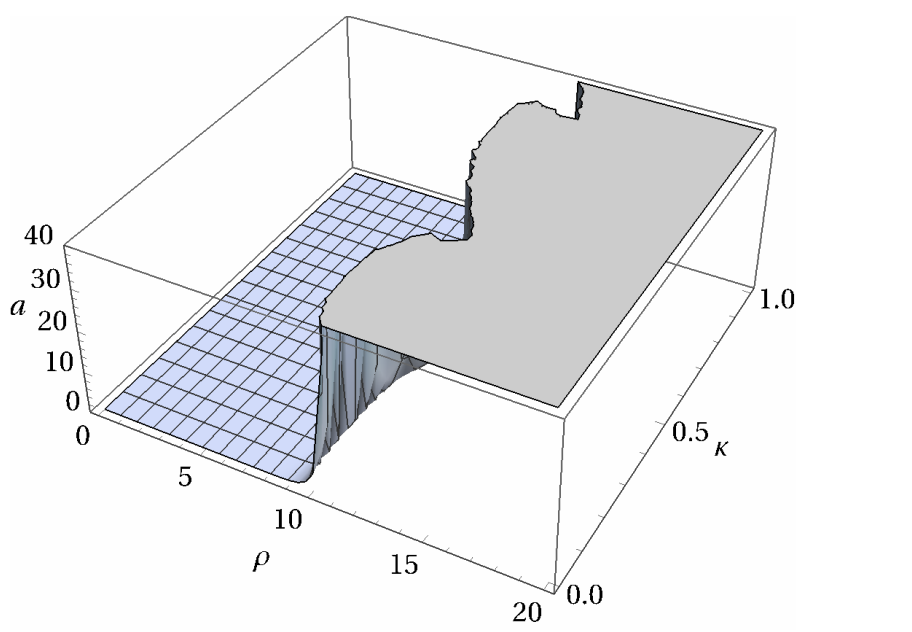}}
\newcommand{\smallrhowlambdaone}{\includegraphics[width=4cm]{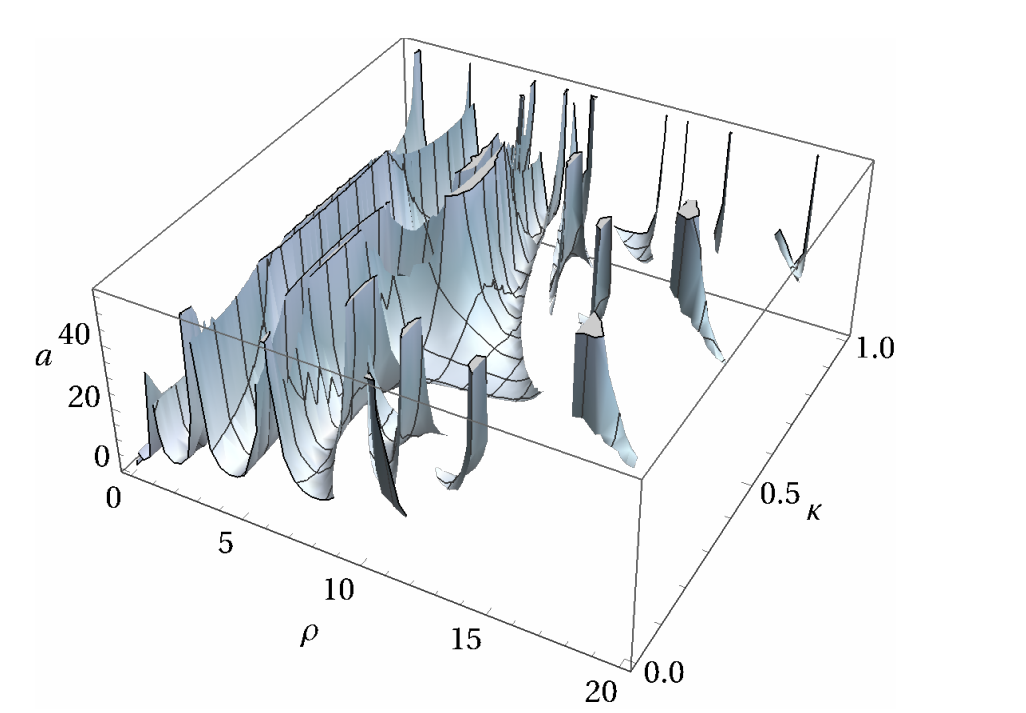}}
\newcommand{\largerhowolambdaone}{\includegraphics[width=4cm]{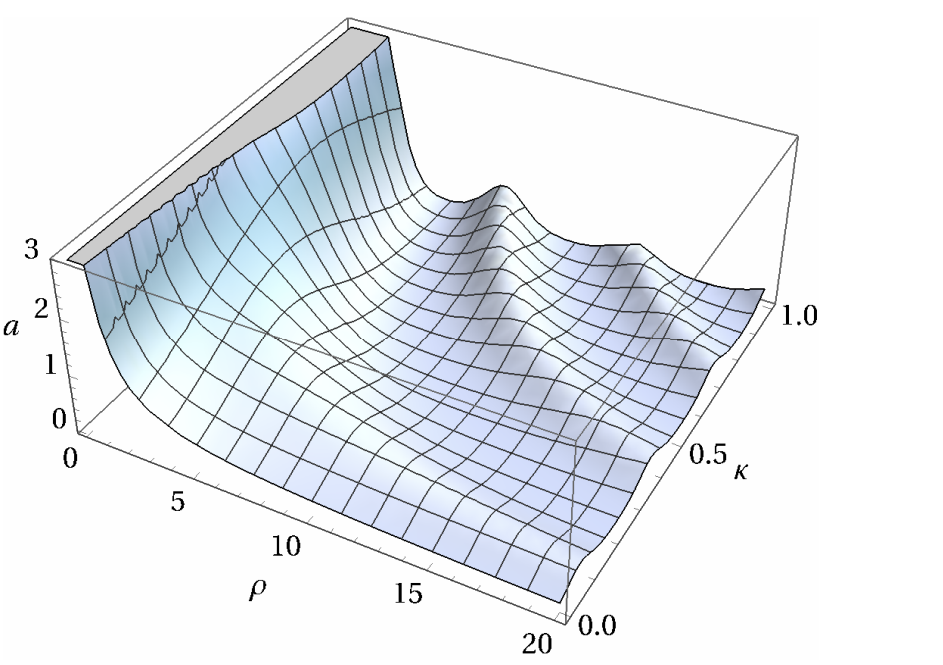}}
\newcommand{\smallrhowolambdaone}{\includegraphics[width=4cm]{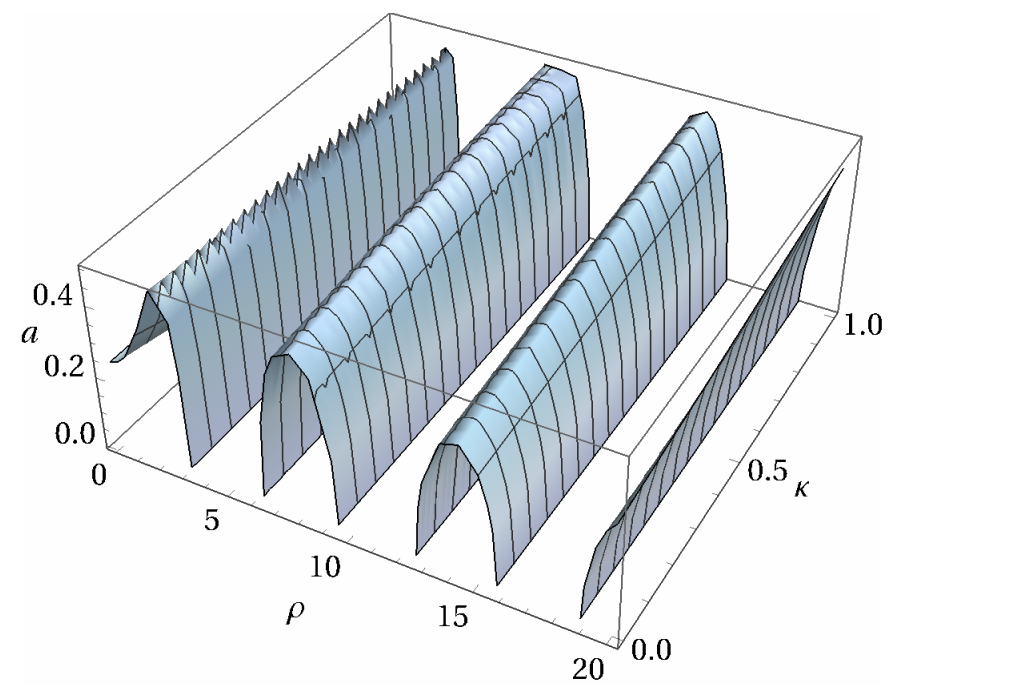}}
\newcommand{\largerhowlambdatwo}{\includegraphics[width=4cm]{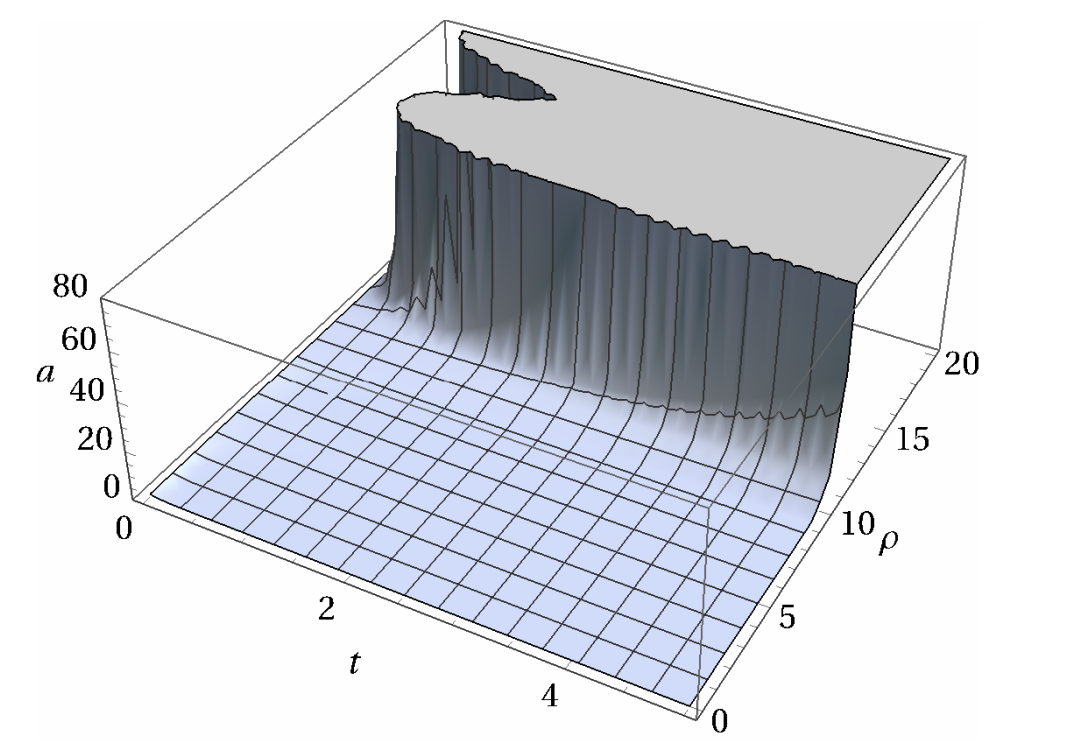}}
\newcommand{\smallrhowlambdatwo}{\includegraphics[width=4cm]{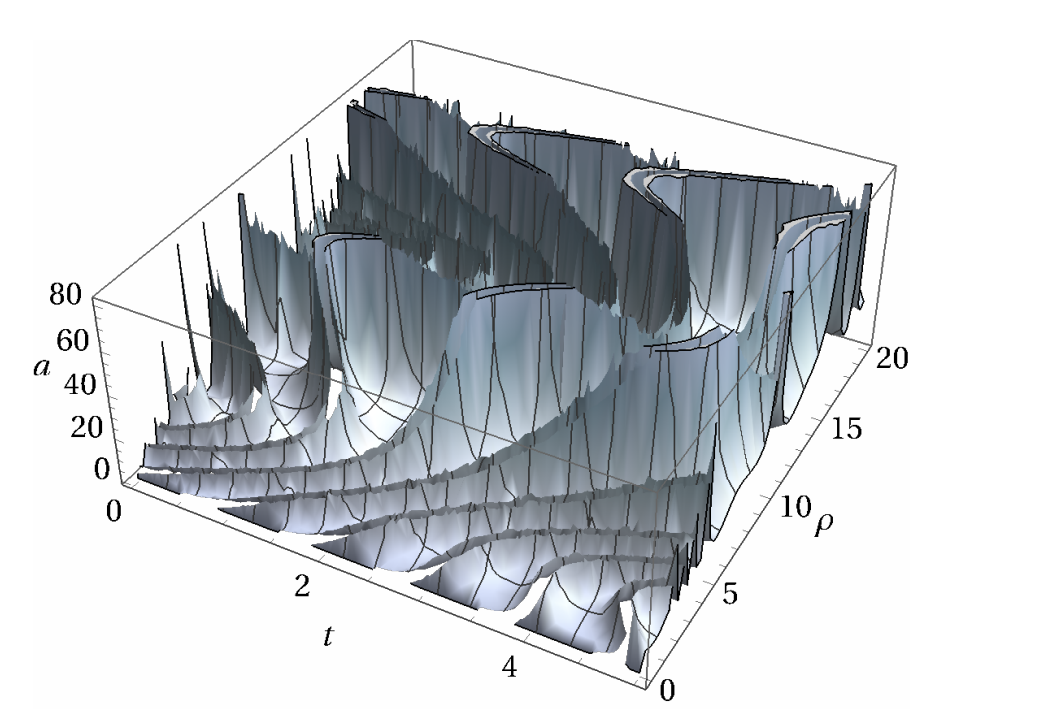}}
\newcommand{\largerhowolambdatwo}{\includegraphics[width=4cm]{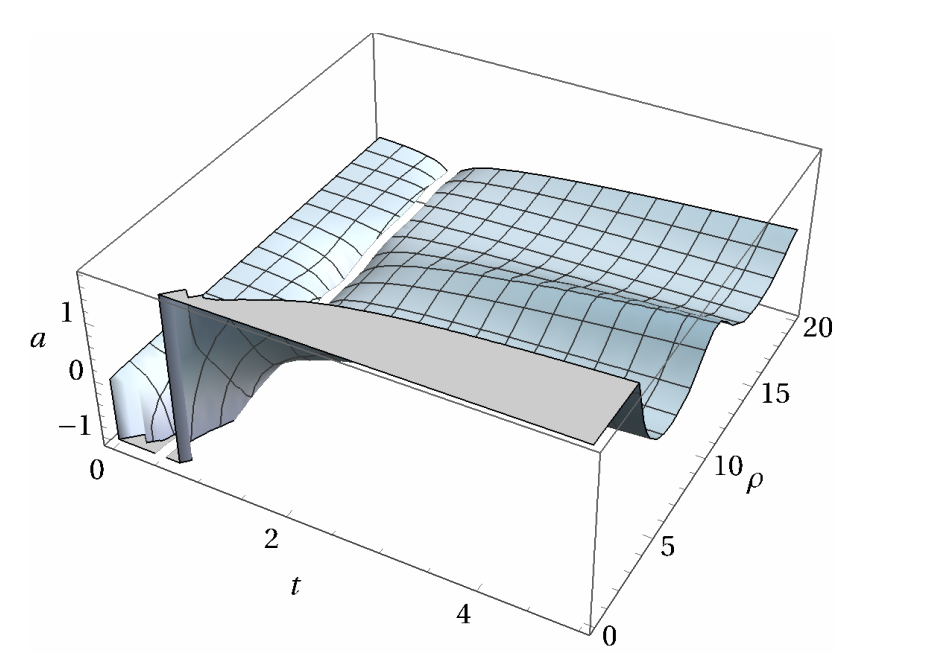}}
\newcommand{\smallrhowolambdatwo}{\includegraphics[width=4cm]{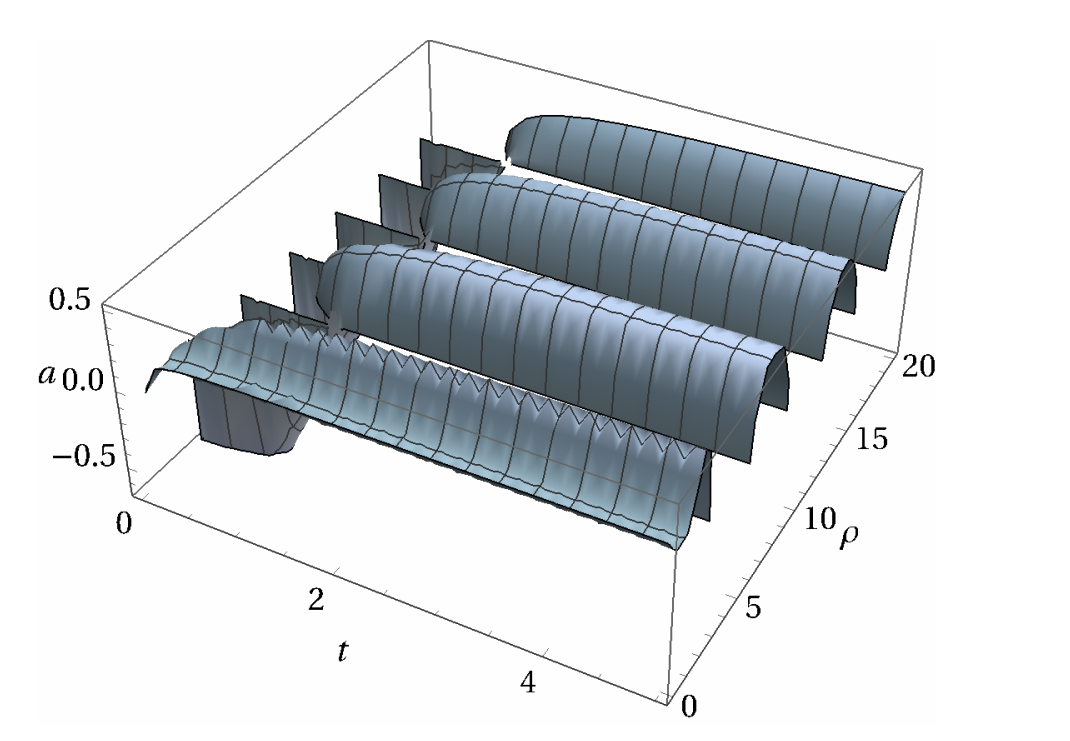}}

\newcolumntype{C}[2]{>{\columncolor{#1}\centering\arraybackslash}m{#2}}

\begin{table}
\centering
    \begin{tabular}{C{white!30}{3em}C{white!20}{5em}C{white!20}{11em}C{white!20}{11em}}
    \toprule
    $\Lambda$ & Parameters & Small $\rho$ & Large $\rho$ \\
    \midrule\\
    Zero    & $(\rho, \kappa,a)$ & \smallrhowolambdaone & \largerhowolambdaone \\ \\
    Zero    & $(t, \rho,a) $     & \smallrhowolambdatwo & \largerhowolambdatwo \\\\
    Nonzero & $(\rho, \kappa,a)$ & \smallrhowlambdaone  & \largerhowlambdaone  \\\\
    Nonzero & $(t, \rho,a)$      & \smallrhowlambdatwo  & \largerhowlambdatwo  \\\\
    \bottomrule
    \end{tabular}
    \caption{These figures are organized based on the absence or presence of dark energy, choice of parameters, and the limit of $\rho$. The  discontinuities and inhomogeneities in the small-$\rho$ limit are sewn together and smoothed out in the large-$\rho$ limit, showing the large-scale cosmological structure. In this picture we have set constant the cylindrical coordinate $z=0.1$ and the free parameters $\alpha=1$ and $\beta=1/2$. We have also set $t=3$ and $\kappa=0.7$ where necessary.}
\end{table}

In both cases our choice of $\Lambda$ results in the smoothening out of the set of deformations in the small-$\rho$ regime as we transition to the large-$\rho$, global, topological regime. The slices plotted can be thought of as being taken off of the cosmic cylinder and flattened to the plane to observe the corresponding topology. Inhomogeneities are also present, reminiscent of the inhomogeneities inherent in the distribution of galaxies at smaller length scales. This crude toy model, in which we have constructed an inherent feature of spacetime that is independent of a particular metric  mimics the global evolutionary features we would expect in transitioning from the local distribution of galaxies to the global homogeneity and isotropy fixed by the cosmological principle. This gives us the bounds to now consider the role of such an inherent structure  given that it is plausible that such a structure could be a mechanism for global evolution. This is motivated by the fact that although our model is, again, highly speculative, modern physical cosmology must account for the very real smoothing of the cosmic manifold that occurs from tracking the distribution of galaxies over large scales. These considerations can be made independent of the mechanism attached to the phenomena described and many have proposed alternative ideas [15]. However, at the time of writing this, all such mechanisms are equally likely.\footnote{One could possible add another component to the stress energy tensor which incorporates the relevant power spectrum

$P(k)=\displaystyle\int d^3r \xi(r)e^{i\vec{k}\cdot\vec{r}}=\frac{<|\delta_{\vec{k}}|^2>}{n^2V_u}-\frac{1}{n}$

sourced by a corresponding $N$-point correlation function.}

How, then are we ought to classify this entity or mechanism? At first glance, we are left first and foremost with a question of scales. The case can be made that when speaking of spacetime as an entity onto itself, or certainly when speaking of a manifold which is most general, we should choose that manifold which describes the largest physical scales. Cosmology, then would be the first place to go. However, it is obvious that we then lose the descriptive power of any system that is not strictly cosmological, like black holes for example. But let us go back to the initial origin of the \emph{ansatz}. The conception of $\Lambda$ was on behalf of the stress energy tensor and not the metric.

In other words, the stuff within the physical system has just as much ontological precedence as the metrical structure  itself and many have argued this detail. This is independent of the notion that our ``stuff" is dynamical and responsible for the shape of the manifold itself. However, this is an important distinction. It is what makes $\Lambda$ in this case inherent to the spacetime structure. The fact that it is dynamical means that we ought to yield it a greater weight within our set of ontological commitments. 

In some sense, if such a function were to exist, its effects would present in a broad class of systems, even if the effects were small. Thinking of the question of scales led us first to the question of applicability in a practical physical context. That such a function's effects are, in principle, present across a larger class of gravitational systems and subsystems means that it should be given greater weight than, say, the $V$ parameter introduced prior. Regardless of metric, the intrinsic shape of the manifold then is shifting adinfinitum albeit in very small, but non-zero, amounts.

 It is conceivable  that $\Lambda$ or the unidentified Dark Matter, which seems more plausible from a phenomenological point of view today, may be ubiquitous across a large enough subset of gravitational phenomena to give it the sense of ontological necessity we have described. At the very least, our ontological commitments must be revised if there is an entity which is intrinsic to the entire manifold upon which separate physical systems can be constructed.

\section{Conclusions} 

In this work, we have explored a set of structures in GR which are linked to the affine structure of the metric and are, at the same time to some degree, independent of it. This new class of intermediary structures, extend the work in [9], and expand that analysis in the context of classical gravitation. We have sought to classify such structures in a particular ontological framework. These structures are given precedence according to the amount of physical data they map onto as well as their practical usage in how one utilizes GR. It would be interesting to explore a dynamical $\Lambda$ which, again, is more aligned with the cosmological data.  It may be an open line of investigation to see if a broader philosophical scheme could be structured which explores the ontology of these entities in greater detail and with greater precision.

\section*{Acknowledgements}
		
 		We are grateful to Peter J. Lewis and Robert Danning for stimulating conversations while completing this work. We also wish to thank P.J.E. Peebles for useful insights and clarifications throughout the course of this work. We also wish to thank Shamik Banerjee and Alok Laddha for helpful clarifications with regards to the determination of falloffs at $\mathcal{I}$. We are also indebted to Erik Curiel for his insights as this work was being completed. 

   \begin{appendix}
	
		\section{Explicit Transformation to Kruskal Coordinates}
		\label{appA}
For those from a pure philosophical background, we explicitly compute the transformation from Schwarzchild to Kruskal coordinates here. We begin with the familiar metric in Schwarzschild coordinates
\begin{equation}
    ds^2=-\left(1-\frac{2M}{r}\right)dt^2+\left(1-\frac{2M}{r}\right)^{-1}dr^2+r^2d\Omega^2
\end{equation}
and impose the following coordinate transformation
\begin{equation}
    \begin{split}
        u=t-r-r_s\ln(r/r_s-1) \\
        v=t+r+r_s\ln(r/r_s-1)
    \end{split}
\end{equation}
Here $r_s$ refers to the Schwarzchild radius. Differentiating these coordinates, one finds that 
\begin{equation}
    \begin{split}
        du=dt-\frac{dr}{1-r_s/r} \\
        dv=dt+\frac{dr}{1-r_s/r}
    \end{split}
\end{equation}
Solving for $dt$ and $dr$ one finds the following expressions
\begin{equation}
\begin{split}
    dt=\frac{1}{2}(dv+du) \\
    dr=\frac{1}{2}(dv-du)(1-r_s/r)
\end{split}
\end{equation}
and the new metric becomes
\begin{equation}
    ds^2=-\left(1-\frac{r_s}{r}\right)dudv+r^2d\Omega^2
\end{equation}
Note that this  is only valid for $r>r_s$. We also have the following relations
\begin{equation}
    \begin{split}
        r+r_s\ln(r/r_s-1)=(v-u)/2 \\
        (r/r_s-1)e^{r/r_s}=e^{(v-u)/2r_s}
    \end{split}
\end{equation}
We want to rewrite the metric (A.5), so using (A.6) we find that
\begin{equation}
    1-\frac{r_s}{r}=\frac{r_s}{r}e^{-r/r_s}e^{\frac{v-u}{2r_s}}
\end{equation}
Plugging this into the metric (A.5) we find the following 
\begin{equation}
    ds^2=-\frac{r}{r_s}e^{-r/r_s}e^{\frac{v-u}{2r_s}}dudv+r^2+d\Omega^2
\end{equation}
We now make a second change of coordinates
\begin{equation}
    \begin{split}
        U=r_se^{-u/2r_s} \\
        V=r_s e^{v/2r_s}
    \end{split}
\end{equation}
Differentiating $U$ and $V$ gives us the following
\begin{equation}
    \begin{split}
        dU=-\frac{1}{2}e^{-u/2r_s}du=-\frac{U}{2r_s}du \\
        dV=\frac{1}{2}e^{v/2r_s}dv=\frac{V}{2r_s}dv
    \end{split}
\end{equation}
and we are left with the following expressions 
\begin{equation}
    \begin{split}
        du=-\frac{2r_s}{U}dU \\
        dv=\frac{2r_s}{V}dV
    \end{split}
\end{equation}
Plugging these into (A.8) gets us the following new metric in terms of $U$ and $V$.
\begin{equation}
    ds^2=\frac{4r_s}{r}e^{-r/r_s}dUdV+r^2d\Omega^2
\end{equation}
For the final transformation, we let $U=X-T$ and $V=X+T$. Then, $dU=-dT+dX$ and $dV=dT+dX$. The resulting metric for $(T,X,\theta,\phi)$ is
\begin{equation}
    ds^2=\frac{4r_s}{r}e^{-r/r_s}(-dT^2+dX^2)+r^2d\Omega^2
\end{equation}
This is what is referred to as the Kruskal metric. As desired, we now have a maximally-extended spacetime with no singularity at $r=r_s$, although the true physical singularity remains. Some additional useful relations to note are
\begin{equation}
    \begin{split}
        X^2-T^2=UV=r_s^2e^{(v-u)/2r_s}=r_s^2e^{r/r_s}(r/r_s-1) \\
        t=(u+v)/2=r_s\ln(V/U)=r_s\ln\left|\frac{X+T}{X-T}\right|
    \end{split}
\end{equation}

\end{appendix}

\end{document}